\documentclass[prb,twocolumn,showpacs,amsmath,amssymb]{revtex4}

\usepackage{graphicx}
\usepackage{dcolumn}
\usepackage{bm}

\begin{document}
\title{Entropy Change through
Rayleigh-B\'enard Convective Transition\\ with Rigid Boundaries}

\author{Takafumi Kita}
\affiliation{Department of Physics, Hokkaido University, Sapporo 060-0810, Japan}

\date{\today}

\begin{abstract}
The previous investigation 
on Rayleigh-B\'enard convection of a dilute classical gas
[T.\ Kita:\ J.\ Phys.\ Soc.\ Jpn.\  {\bf 75} (2006) 124005]
is extended to calculate entropy change of the convective
transition with the rigid boundaries.
We obtain results qualitatively similar to those of
the stress-free boundaries. Above the critical Rayleigh number,
the roll convection is realized
among possible steady states with periodic structures, carrying
the highest entropy as a function of macroscopic mechanical
variables.
\end{abstract}


\maketitle

\section{Introduction}

In a preceding paper,\cite{Kita06b} we performed a statistical mechanical 
investigation on Rayleigh-B\'enard convection of a dilute classical gas 
based on the Boltzmann equation. 
We specifically calculated entropy change through the convective transition 
for the case of the stress-free boundaries as a function of 
macroscopic mechanical variables.
We thereby tested the 
validity of the principle of maximum entropy 
proposed for nonequilibrium steady states.\cite{Kita06a}
The present paper extends the consideration to 
a more realistic case of the rigid boundaries.\cite{Chandrasekhar61}

\section{Formulation and Numerical Procedures}

Our starting point is Eq.\ (46) of ref.\ \onlinecite{Kita06b}:
\begin{subequations}
\label{Boussinesq}
\begin{equation}
\frac{\partial T^{(1)}}{\partial t}+\hat{\bm j}^{(1.5)}\cdot{\bm\nabla}T^{(1)}
-\kappa^{(2)}\nabla^{2}T^{(1)}=0 ,
\end{equation}
\begin{eqnarray}
&&\hspace{-15mm}
-\frac{\partial}{\partial t}\nabla^{2}\hat{\bm j}^{(1.5)}+
{\bm\nabla}\!\times\!{\bm\nabla}\!\times(
\hat{\bm j}^{(1.5)}\!\cdot{\bm\nabla}\hat{\bm j}^{(1.5)})
\nonumber \\
&&\hspace{-15mm}
+\nu^{(2)}(\nabla^{2})^{2}\hat{\bm j}^{(1.5)}
+U_{g}^{(2)}({\bm e}_{z}\nabla^{2}\!-\!{\bm e}_{z}\!\cdot\!{\bm\nabla}{\bm\nabla})
T^{(1)}={\bm 0},
\end{eqnarray}
\end{subequations}
where ${\bm\nabla}\cdot\hat{\bm j}^{(1.5)}\!=\!{\bf 0}$ and the units are described in
\S3.1.
We adopt the condition of the rigid boundaries along $z$, i.e., $\hat{\bm j}^{(1.5)}\!=\! {\bf 0}$
at $z\!=\!\pm 1/2$.
Combined with ${\bm\nabla}\!\cdot\!\hat{\bm j}^{(1.5)}\!=\!{\bf 0}$, it
yields the boundary conditions:
\begin{subequations}
\label{BC}
\begin{equation}
\left.\hat{\bm j}^{(1.5)}_{\perp}\right|_{z =\pm 1/2}\!=\! {\bm 0},
\label{BC1}
\end{equation}
\begin{equation}
\left.\hat{j}^{(1.5)}_{z}\right|_{z =\pm 1/2}\!=\! 
\left.\frac{\partial \hat{j}^{(1.5)}_{z}}{\partial z}\right|_{z =\pm 1/2}=0,
\label{BC2}
\end{equation}
\end{subequations}
with $\hat{\bm j}^{(1.5)}_{\perp}$ denoting the $xy$ components.
As for the horizontal directions,
we consider the region $-L/2\!\leq \! x,y\!\leq\! L/2$
with $L\!\gg\! 1$ and impose the periodic boundary conditions.
We also fix macroscopic mechanical variables of the system,
i.e., the total particle number, energy, and energy flux along $z$.
These conditions lead to eq.\ (42) of ref.\ \onlinecite{Kita06b}, i.e.,
\begin{subequations}
\label{boundary-T}
\begin{equation}
\int T^{(1)}\,{\rm d}^{3}r=0 ,
\label{boundary-T1}
\end{equation}
\begin{equation}
\frac{1}{L^{2}}\int_{-L/2}^{L/2}{\rm d}x
\int_{-L/2}^{L/2}{\rm d}y \left.
\frac{\partial T^{(1)}}{\partial z}\right|_{z=-1/2}=
-\Delta T_{\rm hc} ,
\label{boundary-T2}
\end{equation}
\end{subequations}
where $\Delta T_{\rm hc}\!
\equiv\!{2\bar{j}_{Q}^{(3)}}/{3\bar{n}\kappa^{(2)}}$ denotes
the temperature difference between $z\!=\!\pm 1/2$ that would be realized
in the heat-conducting state.

We solve the above equations with the method
developed by Pesch.\cite{Pesch96}
First, the boundary conditions of eq.\ (\ref{BC}) is treated with the Galerkin method,\cite{Galerkin}
i.e., by expanding every $z$ dependence in terms of some basis functions
satisfying the boundary conditions.
Specifically, the basis functions for eq.\ (\ref{BC1}) are
obtained from the second-order differential equation $S''=-\lambda S$ with $S(\pm 1/2)\!=\! 0$ as
\begin{subequations}
\label{bases}
\begin{equation}
S_{n}(z)=\sqrt{2}\sin n\pi(z\!+\! 1/2) \hspace{5mm} (n=1,2,\cdots).
\end{equation}
They satisfy $\langle S_{n}|S_{n'}\rangle\!\equiv\!\int_{-1/2}^{1/2}S_{n}(z)S_{n'}(z){\rm d}z
\!=\!\delta_{nn'}$.
On the other hand, those for eq.\ (\ref{BC2}) are constructed from
the fourth-order differential equation
$C^{(4)}=k^{4}C$ with
$C(\pm 1/2)\!=\!C'(\pm 1/2)\!=\! 0$ as
\begin{equation}
C_{n}(z)=
\left\{
\begin{array}{ll}
\vspace{2mm}
\!\!\displaystyle
A_{n}\!\left(\frac{\cosh k_{n}z}{\cosh\frac{k_{n}}{2}}-
\frac{\cos k_{n}z}{\cos\frac{k_{n}}{2}}\right)
& : n=1,3,\cdots \\
\displaystyle
\!\!A_{n}\!\left(\frac{\sinh k_{n}z}{\sinh\frac{k_{n}}{2}}-
\frac{\sin k_{n}z}{\sin\frac{k_{n}}{2}}\right)
& : n=2,4,\cdots 
\end{array}
\right. ,
\end{equation}
where $k_{n}(>\!0)$ is determined by
\begin{equation}
\left\{
\begin{array}{ll}
\vspace{2mm}
\!\!\displaystyle
\tanh\frac{k_{n}}{2}+\tan\frac{k_{n}}{2}=0
& : n=1,3,\cdots \\
\!\!\displaystyle
\coth\frac{k_{n}}{2}-\cot\frac{k_{n}}{2}=0
& : n=2,4,\cdots 
\end{array}
\right. .
\end{equation}
Note $k_{n}\!\approx\! (n\!+\!1/2)\pi$ for $n\! \gg\! 1$.
The quantity $A_{n}$ is the normalization constant:
\begin{equation}
A_{n}=\left\{
\begin{array}{ll}
\vspace{3mm}
\displaystyle
\frac{\sqrt{2}\cosh\frac{k_{n}}{2}\cos\frac{k_{n}}{2}}
{\sqrt{\cosh^{2}\frac{k_{n}}{2}+\cos^{2}\frac{k_{n}}{2}}}
& :n=1,3,\cdots\\
\displaystyle
\frac{\sqrt{2}\sinh\frac{k_{n}}{2}\sin\frac{k_{n}}{2}}
{\sqrt{\sinh^{2}\frac{k_{n}}{2}-\sin^{2}\frac{k_{n}}{2}}}
 & :n=2,4,\cdots
\end{array}\right. ,
\label{An}
\end{equation}
\end{subequations}
so that $\langle C_{n}|C_{n'}\rangle
\!=\!\delta_{nn'}$.
The functions $\{C_{n}(z)\}$ may be called Chandrasekhar functions.\cite{Chandrasekhar61}

Now that appropriate basis functions are obtained, 
we expand $T^{(1)}$ and $\hat{\bm j}^{(1.5)}$ in eq.\ (\ref{Boussinesq}) as
\begin{subequations}
\label{expansions}
\begin{equation}
T^{(1)}({\bm r})=-\Delta T z -T_{1}+\sum_{{\bm k}_{\perp}}\sum_{n=1}^{\infty}
\tilde{T}({\bm k}_{\perp},n)
{\rm e}^{i{\bm k}_{\perp}\cdot{\bm r}}S_{n}(z),
\label{expansionsT}
\end{equation}
\begin{equation}
\hat{j}^{(1.5)}_{z}({\bm r})=\sum_{{\bm k}_{\perp}\neq {\bf 0} }\sum_{n=1}^{\infty}
\tilde{j}_{z}({\bm k}_{\perp},n)
{\rm e}^{i{\bm k}_{\perp}\cdot{\bm r}}C_{n}(z),
\label{expansionsZ}
\end{equation}
\begin{equation}
\hat{\bm j}^{(1.5)}_{\perp}({\bm r})=\sum_{{\bm k}_{\perp}\neq {\bf 0}}\sum_{n=1}^{\infty}
\tilde{\bm j}_{\perp}({\bm k}_{\perp},n)
{\rm e}^{i{\bm k}_{\perp}\cdot{\bm r}}S_{n}(z).
\label{expansionsP}
\end{equation}
\end{subequations}
Here ${\bm k}_{\perp}\!=\!{\bf 0}$ component is excluded in the expansion of $\hat{\bm j}^{(1.5)}$
to seek only periodic current distributions in the $xy$ plane.
We have also incorporated into eq.\ (\ref{expansionsT}) 
the fact that the temperature is uniform at $z\!=\!\pm 1/2$.
Now, ${\bm\nabla}\cdot\hat{\bm j}^{(1.5)}\!=\!{\bf 0}$ is transformed into
$i{\bm k}_{\perp}\cdot\tilde{\bm j}_{\perp}({\bm k}_{\perp},n)
\!+\!\sum_{n'}\langle S_{n} |
C_{n'}'\rangle \tilde{j}_{z}({\bm k}_{\perp},n')\!=\!0$.
It hence follows that $\tilde{\bm j}_{\perp}({\bm k}_{\perp},n)$
can be expressed generally as
\begin{eqnarray}
&&\hspace{-12mm}
\tilde{\bm j}_{\perp}({\bm k}_{\perp},n)
\nonumber \\
&&\hspace{-12mm}
=i\frac{{\bm k}_{\perp}}{k_{\perp}^{2}}\sum_{n'}\langle S_{n} |
C_{n'}'\rangle \tilde{j}_{z}({\bm k}_{\perp},n')+\frac{{\bm e}_{z}\times
{\bm k}_{\perp}}{k_{\perp}}\tilde{j}_{p}({\bm k}_{\perp},n) .
\label{j_perp}
\end{eqnarray}
On the other hand, eq.\ (\ref{boundary-T})
is transformed into
\begin{subequations}
\label{T1dT}
\begin{equation}
T_{1} = \sum_{m=1}^{\infty}
\frac{2\sqrt{2}\,\tilde{T}({\bm 0},2m-1)}{(2m\!-\!1)\pi},
\label{T1}
\end{equation}
\begin{equation}
\Delta T = \Delta T_{\rm hc}+\sum_{n=1}^{\infty}\sqrt{2} n\pi \tilde{T}({\bm 0},n).
\label{dT}
\end{equation}
\end{subequations}
Let us substitute eq.\ (\ref{expansions}) with eq.\ (\ref{j_perp})
into eq.\ (\ref{Boussinesq}) and perform space integrations
using the orthonormality of the basis functions.
We thereby obtain algebraic equations for the expansion coefficients as
\begin{subequations}
\label{Boussinesq-Alg}
\begin{eqnarray}
&&\hspace{-10mm}
\frac{\partial \tilde{T}({\bm k}_{\perp},n)}{\partial t}
+\kappa^{(2)}[k_{\perp}^{2}\!+\!(n\pi)^{2}]\tilde{T}
({\bm k}_{\perp},n)
\nonumber \\
&&\hspace{-10mm}
-\Delta T \sum_{n'}
\langle S_{n}|C_{n'}\rangle \tilde{j}_{z}({\bm k}_{\perp},n')
\nonumber \\
&&\hspace{-10mm}
=
-\frac{1}{L^{2}}\langle {\rm e}^{i{\bm k}_{\perp}\cdot{\bm r}}S_{n} |
\hat{\bm j}^{(1.5)}\!\cdot{\bm\nabla}\tilde{T}^{(1)}\rangle ,
\label{Boussinesq-Alg1}
\end{eqnarray}
\begin{eqnarray}
&&\hspace{-10mm}
\sum_{n'}
(k_{\perp}^{2}\delta_{nn'}\!-\langle C_{n}|C_{n'}''\rangle)
\frac{\partial \tilde{j}_{z}({\bm k}_{\perp},n')}{\partial t}
\nonumber \\
&&\hspace{-10mm}
+\nu^{(2)}\sum_{n'}\bigl[
(k_{\perp}^{4}\!+\!k_{n}^{4})^{2}\delta_{nn'}-2k_{\perp}^{2}
\langle C_{n}|C_{n'}''\rangle \bigr] \tilde{j}_{z}
({\bm k}_{\perp},n')
\nonumber \\
&&\hspace{-10mm}
-U_{g}^{(2)}k_{\perp}^{2}\sum_{n'}
\langle C_{n} |S_{n'}\rangle \tilde{T}({\bm k}_{\perp},n')
\nonumber \\
&&\hspace{-10mm}
=
-\frac{1}{L^{2}}{\bm e}_{z}\cdot \langle {\rm e}^{i{\bm k}_{\perp}\cdot{\bm r}}C_{n} |
{\bm\nabla}\!\times\!{\bm\nabla}\!\times(
\hat{\bm j}^{(1.5)}\!\cdot{\bm\nabla}\hat{\bm j}^{(1.5)})\rangle,
\label{Boussinesq-Alg2}
\end{eqnarray}
\begin{eqnarray}
&&\hspace{-10mm}
[k_{\perp}^{2}\!+\!(n\pi)^{2}]\frac{\partial \tilde{j}_{p}
({\bm k}_{\perp},n)}{\partial t}+\nu^{(2)}
[k_{\perp}^{2}\!+\!(n\pi)^{2}]^{2}\tilde{j}_{p}
({\bm k}_{\perp},n)
\nonumber \\
&&\hspace{-10mm}=
-\frac{1}{L^{2}}\frac{
{\bm e}_{z}\times {\bm k}_{\perp}}{k_{\perp}}\cdot
\langle {\rm e}^{i{\bm k}_{\perp}\cdot{\bm r}}S_{n} |
{\bm\nabla}\!\times\!{\bm\nabla}\!\times(
\hat{\bm j}^{(1.5)}\!\cdot{\bm\nabla}\hat{\bm j}^{(1.5)})\rangle,
\nonumber \\
\label{Boussinesq-Alg3}
\end{eqnarray}
\end{subequations}
with $\langle f |g\rangle\!\equiv\!\int_{-L/2}^{L/2}{\rm d}x\int_{-L/2}^{L/2}{\rm d}y
\int_{-1/2}^{1/2}{\rm d}z f^{*}({\bm r})g({\bm r})$.

Finally, entropy characteristic of convection is obtained 
by substituting eq.\ (\ref{expansionsT})
into eq.\ (50) of ref.\ \onlinecite{Kita06b} as
\begin{eqnarray}
&&\hspace{-10mm}
S^{(2)}
=-\frac{5}{4}\biggl\{-T_{1}^{2}+\frac{(\Delta T)^{2}}{12}
+\sum_{{\bf k}_{\perp}}\sum_{n=1}^{\infty} |\tilde{T}({\bf k}_{\perp},n)|^{2}
\nonumber \\
&&\hspace{8mm}
+\Delta T \sum_{m=1}^{\infty} 
\frac{\sqrt{2}\tilde{T}({\bf 0},2m)}{m\pi} 
\biggr\},
\label{S^2}
\end{eqnarray}
with $T_1$ and $\Delta T$ given by eq.\ (\ref{T1dT}).

It follows from the stability analysis for the heat-conducting state\cite{Chandrasekhar61} 
that the critical Rayleigh number $R_{\rm c}$ is determined 
from eq.\ (\ref{Boussinesq-Alg}) 
by setting the nonlinear terms and time derivatives 
equal to zero.
The relevant instability originates from the linear coupled equations
for $\tilde{T}({\bm k},n)$ and $\tilde{j}_{z}({\bm k},n)$.
Eliminating $\tilde{T}({\bm k},n)$ in favor of $\tilde{j}_{z}({\bm k},n)$,
we obtain the equation for $R_{\rm c}$ as
\begin{equation}
\det \underline{A}=0,
\label{detA}
\end{equation}
where matrix $\underline{A}$ is defined by
\begin{eqnarray}
&&\hspace{-12mm}
A_{nn'}\equiv
(k_{\perp}^{4}\!+\!k_{n}^{4})^{2}\delta_{nn'}-2k_{\perp}^{2}
\langle C_{n}|C_{n'}''\rangle 
\nonumber \\
&&\hspace{-0mm}
-R^{(-1)}k_{\perp}^{2}\sum_{n''}\frac{
\langle C_{n} |S_{n''}\rangle\langle S_{n''} |C_{n'}\rangle }{k_{\perp}^{2}
+(n''\pi)^{2}},
\label{Eq-Rc0}
\end{eqnarray}
with
\begin{equation}
R^{(-1)}\equiv
\frac{U_{g}^{(2)}\Delta T}{\nu^{(2)}\kappa^{(2)}}.
\end{equation}
The critical Rayleigh number $R_{\rm c}$ corresponds to the minimum value of $R^{(-1)}$ 
in eq.\ (\ref{detA}) as a function of $k_{\perp}$.
Equation (\ref{detA}) is solved by approximating $\underline{A}$ by a finite dimension of
$n_{\rm c}\!\times\! n_{\rm c}$, and the convergence is checked by increasing $n_{\rm c}$.
Choosing $n_{\rm c}=4$ already yields an excellent result of
$R_{\rm c}=1.708$ with $k_{\perp}=3.116\equiv k_{\rm c}$.\cite{Chandrasekhar61}

The nonlinear terms become
relevant in eq.\ (\ref{Boussinesq-Alg}) for $R^{(-1)}\!>\! R_{\rm c}$. 
They are evaluated for given expansion
coefficients $\tilde{T}({\bm k}_{\perp},n)$ and $\tilde{\bm j}({\bm k}_{\perp},n)$ as follows.
We first construct $T^{(1)}({\bm r})$ and $\hat{\bm j}^{(1.5)}({\bm r})$
by eqs.\ (\ref{expansions})-(\ref{T1dT}). 
The fast Fourier transform (FFT)\cite{NR} is used in this procedure to obtain the $xy$ dependence.
We then perform the space differentiations numerically
in the $xy$ plane and analytically along the $z$ direction. 
We specifically use the following formulas of $O(h^{6})$ in the $xy$ plane:
\begin{subequations}
\label{Nd}
\begin{eqnarray}
&&\hspace{-10mm}
\partial_{x} f(x,y)\approx \sum_{\sigma=\pm }\frac{\sigma}{60h}
[f(x\!+\!3\sigma h,y)
\!-\!9f(x\!+\!2\sigma h,y)
\nonumber \\
&&\hspace{8mm}
+ 45f(x\!+\!\sigma h,y)],
\label{Nd1}
\end{eqnarray}
\begin{eqnarray}
&&\hspace{-8mm}
\partial_{x}^{2} f(x,y)\approx \frac{1}{180h^{2}}\biggl\{
\sum_{\sigma=\pm }[2f(x\!+\!3\sigma h,y)
\!-\!27f(x\!+\!2\sigma h,y)
\nonumber \\
&&\hspace{10mm}
+ 270f(x\!+\!\sigma h,y)]-490f(x,y)\biggr\}.
\label{Nd2}
\end{eqnarray}
\end{subequations}
The quantity $\partial_{xy} f(x,y)$ is obtained with eq.\ (\ref{Nd1}) 
by averaging the derivatives performed in different order.
The nonlinear overlap integrals are evaluated finally,
where we again use the FFT in the $xy$ plane.
On the other hand, all the calculations along $z$ are performed by preparing 
the relevant overlap integrals in advance, e.g., $\langle S_{n}|S_{n'}S_{n''}\rangle$ and
$\langle S_{n}|C_{n'}S_{n''}'\rangle$ for eq.\ (\ref{Boussinesq-Alg1}),
and performing the summations over $n'$ and $n''$.

Time evolutions of eq.\ (\ref{Boussinesq-Alg}) are calculated as follows.
We first multiply eqs.\ (\ref{Boussinesq-Alg2}) and (\ref{Boussinesq-Alg3}) 
by $(\underline{\cal O}^{-1})_{n''n}$ and $[k_{\perp}^{2}\!+\!(n\pi)^{2}]^{-1}$
with $(\underline{\cal O})_{nn'}\!\equiv\!k_{\perp}^{2}\delta_{nn'}\!-\!\langle C_{n}|C_{n'}''\rangle$,
respectively,
and perform summation over $n$ for eq.\ (\ref{Boussinesq-Alg2}).
Time integrations are then carried out numerically by treating
$\partial {\bm f}/\partial t\!=\! {\bm g}$ as 
${\bm f}(t\!+\!\Delta t)\!\approx\! {\bm f}(t)\!+\!  {\bm g}(t)\Delta t $.
A disadvantage of this simple method is that we have to make $\Delta t $ small enough
to avoid an explosion in the numerical time integration.
One may alternatively use
the split-step integration scheme developed by Pesch which approximates 
$\partial {\bm f}/\partial t\!=\! \underline{L} {\bm f}\!+\! {\bm g}$ 
as ${\bm f}(t\!+\!\Delta t)\!\approx\! {\rm e}^{\,\underline{L} \Delta t} {\bm f}(t)+  
{\rm e}^{\,\underline{L} \Delta t/2} [3{\bm g}(t)- {\bm g}(t\!-\!\Delta t)]\Delta t/2$, thereby
treating the linear part $\underline{L}{\bm f}$ exactly.
This latter scheme removes the explosion at the expense of larger numerical errors
to make a rapid time integration possible; the extra computational time for 
diagonalizing $\underline{L}$ in the calculation of ${\rm e}^{\,\underline{L} \Delta t}$ 
is negligible in the whole numerical procedures.

We here focus on periodic solutions of eq.\ (\ref{Boussinesq}) in the $xy$ plane and 
express ${\bm r}_{\perp}\!=\! s_{1}{\bm a}_{1}\!+\!s_{2}{\bm a}_{2}$, where
${\bm a}_{1}\!\equiv\!(a_{1x},a_{1y},0)$ and ${\bm a}_{2}\!\equiv\!(0,a_{2},0)$ denote the basic
vectors.
Accordingly, we adopt the periodic boundary condition for
the region spanned by ${\cal N}_{1}{\bm a}_{1}$
and ${\cal N}_{2}{\bm a}_{2}$ with ${\cal N}_{j}$ ($j\!=\!1,2$) a large integer.
The above theoretical framework
can also be used in this case with a minor modification. 
Indeed, we only have to perform the change of variables
$(x,y)\!\rightarrow\!(s_{1},s_{2})$ in the $xy$ integrations of the nonlinear terms.
Those integrations have to be carried out now only over the unit cell of $0\!\leq\! s_{1},s_{2}\!\leq\! 1$.
The corresponding wave vector ${\bm k}_{\perp}$ is given by
${\bm k}_{\perp}=\ell_{1}{\bm b}_{1}+\ell_{2}{\bm b}_{2}$, 
where ${\bm b}_{1}\equiv 2\pi({\bm a}_{2}\times{\bm e}_{z})/[({\bm a}_{1}\times{\bm a}_{2})
\cdot{\bm e}_{z}]$,
${\bm b}_{2}\equiv 2\pi({\bm e}_{z}\times{\bm a}_{1})/[({\bm a}_{1}\times{\bm a}_{2})\cdot{\bm e}_{z}]$,
and $\ell_{j}$ denotes an integer.
The linear stability analysis for the heat-conducting state suggests that
the stable solution satisfies
$|{\bm b}_{1}|\!\sim \!|{\bm b}_{2}|\!\sim \! k_{{\rm c}}\!=\!3.116$.

The parameters in eq.\ (\ref{Boussinesq-Alg}) are chosen the same as
those used for the free boundaries, i.e., eq.\ (37a) of ref.\ \onlinecite{Kita06b},
which correspond to Ar at 273K under atmospheric pressure.
We also fix the heat-flux density $\bar{j}_{Q}^{(3)}$ at
$z\!=\! -1/2$ so that the temperature difference $\Delta T_{\rm hc}\!=\! 1$K is realized
between $z\!=\!\pm 1/2$ in the heat-conducting state.
The Rayleigh number $R^{(-1)}$ is controlled by changing the thickness $d$.

Practical calculations of eq.\ (\ref{Boussinesq-Alg}) are performed as follows:
We first multiply eqs.\ (\ref{Boussinesq-Alg1})-(\ref{Boussinesq-Alg3}) 
by $10^{8}$, $10^{10}$ and $10^{10}$, respectively, 
and rewrite them in terms of 
$\tilde{T}'({\bm k}_{\perp},n)\!\equiv\!10^{3}\tilde{T}({\bm k}_{\perp},n)$
and $\tilde{j}'({\bm k}_{\perp},n)\!\equiv\!10^{5}\tilde{j}({\bm k}_{\perp},n)$
to obtain equations of $O(1)$.
The summations over $n$
are truncated at a finite value $n_{\rm c}$,
whereas $N_{\rm FFT}$ discrete points are used to perform FFT
for each direction in the $xy$ plane.
As for periodic structures, we investigate the three 
candidates: the roll, the square lattice
and the hexagonal lattice with $|{\bm b}_{1}|\!= \!|{\bm b}_{2}|\!\sim \! k_{\rm c}$.
We then trace time evolutions of 
the expansion coefficients until they all acquire constant values.
Choosing $\Delta t\!\lesssim \! 0.005$, $n_{\rm c}\!\gtrsim\! 4$ and $N_{\rm FFT}\!\gtrsim\!2^4$
yields excellent convergence for the calculations presented below
even with the simplest time-integration scheme.
The initial state is chosen as the conducting state 
with small fluctuations $\tilde{T}'({\bm k}_{\perp},1)\!\sim\! 10^{-2}$ 
for the basic harmonics ${\bm k}_{\perp}$.
The constants $\Delta T$ and $T_{1}$ are updated at
each time step by using eq.\ (\ref{T1dT}).
Also evaluated at each time step is
entropy measured with respect to the heat-conducting state:
\begin{equation}
\Delta S\equiv S^{(2)}- S^{(2)}_{\rm hc}\, ,
\label{dS}
\end{equation}
where $S^{(2)}$ is given by eq.\ (\ref{S^2}) and $S^{(2)}_{\rm hc}\!=\!
-5(\Delta T_{\rm hc})^{2}/48$.
We thereby trace time evolution of $\Delta S$ simultaneously.
The above procedure is carried out for each fixed periodic structure.

One of the advantages of the present approach is that we only have to change the basis functions
to study other boundary conditions.
For example, the case of the stress-free boundaries can also be treated within
the present framework by simply changing $C_{n}(z)\!\rightarrow\!S_{n}(z)$ and
$S_{n}(z)\!\rightarrow\!\sqrt{2}\cos n\pi(z\!+\!1/2)$ in the expansions of 
eqs.\ (\ref{expansionsZ}) and (\ref{expansionsP}), respectively. 
This replacement has been checked to reproduce the results obtained in ref.\ \onlinecite{Kita06b}
appropriately.

\begin{figure}[t]
\begin{center}
  \includegraphics[width=0.85\linewidth]{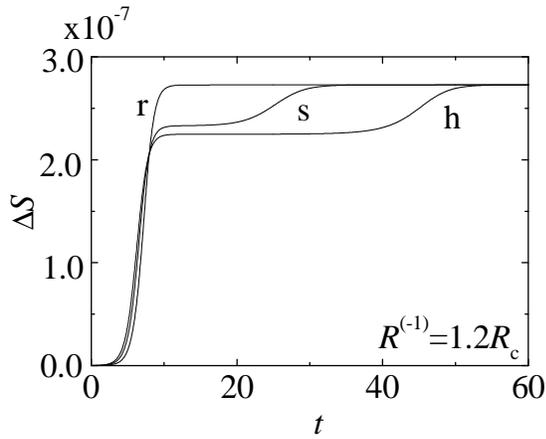}
\end{center}
  \caption{Time evolution of entropy measured with respect to the
  heat-conducting state for $R^{(-1)}\!=\! 1.2R_{\rm c}$. The letters r, s and h denote 
  roll, square and hexagonal, respectively, distinguishing initial fluctuations around
  the heat-conducting solution;
  see text for details. The final state of $t\!\gtrsim\!50$ is the roll convection,
  whereas the intermediate plateaus of s and h correspond to the square and hexagonal
  convections, respectively.}
  \label{fig:1}
\end{figure}

\begin{figure}[t]
\begin{center}
  \includegraphics[width=0.9\linewidth]{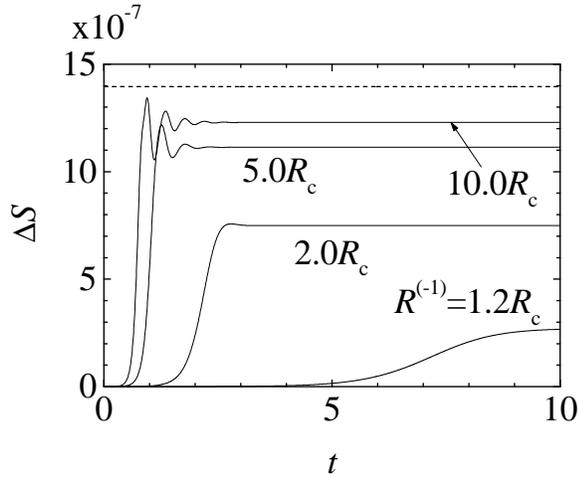}
\end{center}
  \caption{Time evolution of entropy $\Delta S$. 
  The four curves correspond to the different Rayleigh numbers: $R^{(-1)}
  =1.2R_{\rm c}$, $2.0R_{\rm c}$, $5.0R_{\rm c}$ and $10.0R_{\rm c}$.
  The initial state is the
  heat-conducting state with the fluctuation $\tilde{T}'[\pm 1,0,1]\!=\!
  1.0\times 10^{-2}/\sqrt{8}$ and $|{\bm b}_{1}|=k_{\rm c}$, whereas
  all the final states are the roll convection.
  The broken line near the top indicates the upper bound of $\Delta S$.}
  \label{fig:2}
\end{figure}

\section{Results}
We now present numerical results on the rigid boundaries,
which turn out to be qualitatively the same as those of the stress-free boundaries.\cite{Kita06b}

Figure \ref{fig:1} shows time evolution of $\Delta S$
for the Rayleigh number $R^{(-1)}\!=\! 1.2R_{\rm c}$.
The letters r, s and h denote (r) roll, (s) square and (h) hexagonal,
respectively, distinguishing initial conditions; they are exactly the same as those
for the stress-free boundaries.\cite{Kita06b}
Writing $\tilde{T}'({\bm k}_{\perp},n)\!=\!\tilde{T}'[\ell_{1},\ell_{2},n]$
and introducing the angle $\theta$ by 
$\theta\!\equiv\cos^{-1}({\bm b}_{1}\cdot{\bf b}_{2})$,
those initial conditions are given explicitly as follows: 
(r) $\tilde{T}'[\pm 1,0,1]\!=\!1.00\times 10^{-2}/\sqrt{8}$ with 
$|{\bm b}_{1}|\!=\!k_{\rm c}$;
(s) $\tilde{T}'[\pm 1,0,1]\!=\!1.01\times 10^{-2}/\sqrt{8}$ and 
$\tilde{T}'[0,\pm 1,1]\!=\!0.99\times 10^{-2}/\sqrt{8}$ 
with $|{\bm b}_{1}|\!=\!|{\bm b}_{2}|\!=\!k_{\rm c}$ and $\theta\!=\!\pi/2$;
(h) $\tilde{T}'[\pm 1,0,1]\!=\!\tilde{T}'[0,\pm 1,1]\!=\!1.00\times 10^{-2}/\sqrt{8}$ and 
$\tilde{T}'[1,1,1]\!=\!\tilde{T}'[-1,-1,1]\!=\!1.01\times 10^{-2}/\sqrt{8}$ 
with $|{\bm b}_{1}|\!=\!
|{\bm b}_{2}|\!=\!k_{\rm c}$ and $\theta\!=\!2\pi/3$.
We observe clearly that entropy increases
monotonically in all the three cases to reach
a common final value of the roll convection.
The intermediate plateaus seen in s and h
correspond to the metastable square and hexagonal lattices,
respectively, with (s) $\tilde{T}'[\pm 1,0,1]\!\sim\!\tilde{T}'[0,\pm 1,1]$
and (h) $\tilde{T}'[\pm 1,0,1]\!\sim\!\tilde{T}'[0,\pm 1,1]\!\sim\!\tilde{T}'[\pm 1,\pm 1,1]$.

Figure \ref{fig:2} displays time evolution of $\Delta S$ for 
four different Rayleigh numbers, all developing from the 
initial fluctuation
$\tilde{T}'[\pm 1,0,1]\!=\!1.00\times 10^{-2}/\sqrt{8}$ with 
$|{\bm b}_{1}|\!=\!k_{\rm c}$.
Each final state is the roll convection.
Compared with the case of the stress-free boundaries,\cite{Kita06b} we observe an enhanced 
oscillatory behavior for $R\!=\! 5.0R_{\rm c}$ and $10.0R_{\rm c}$
after the first rapid increase of $\Delta S$.

\begin{figure}[t]
\begin{center}
  \includegraphics[width=0.9\linewidth]{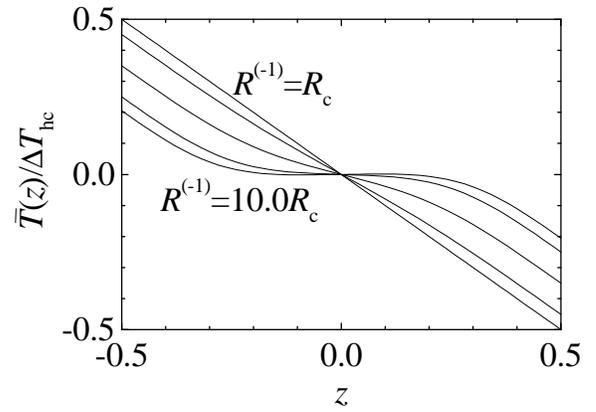}
\end{center}
  \caption{Profile of the average temperature variation $\bar{T}(z)$
  in the roll convection normalized by the temperature difference 
  $\Delta T_{\rm hc}$ in the heat-conducting state. The Rayleigh numbers are
  $R^{(-1)}\!=\! R_{\rm c}$, $1.2R_{\rm c}$, $2.0R_{\rm c}$, $5.0R_{\rm c}$
  and $10.0R_{\rm c}$ from top to bottom on the left part.}
  \label{fig:3}
\end{figure}

\begin{figure}[t]
\begin{center}
  \includegraphics[width=0.9\linewidth]{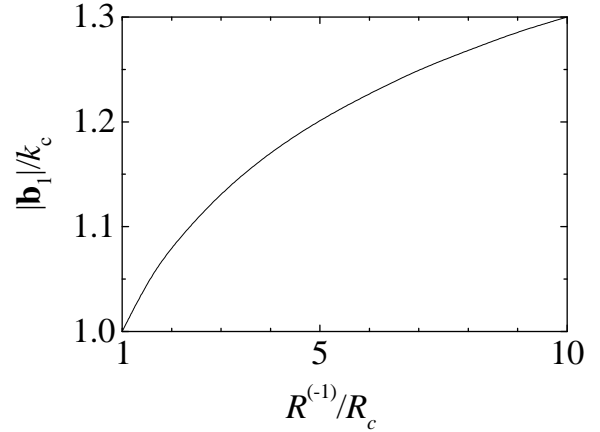}
\end{center}
  \caption{
 The length $|{\bm b}_{1}|/k_{\rm c}$ of the stable roll convection
 as a function of the normalized Rayleigh number $R^{(-1)}/R_{\rm c}$.}
  \label{fig:4}
\end{figure}

Figure \ref{fig:3} shows profile of the average temperature variation
$\bar{T}(z)$ along $z$ in the roll convection for five different
Rayleigh numbers. 
Figure \ref{fig:4} plots $|{\bm b}_{1}|/k_{{\rm c}}$
corresponding to the maximum of $\Delta S$
as a function of $R^{(-1)}/R_{\rm c}$. 
Again the basic features are qualitatively the same as those of the 
stress-free boundaries.\cite{Kita06b}

Thus, we have seen that the principle of 
maximum entropy proposed in ref.\ \onlinecite{Kita06a} is satisfied through 
the Rayleigh-B\'enard convective 
transition of a dilute classical gas
even in the realistic case of the rigid boundaries.

\begin{acknowledgements}
I would like to thank W. Pesch for explaining his 
method to solve the Boussinesq equations.
This work is supported in part by the 21st century COE program 
``Topological Science and  Technology,'' Hokkaido University.
\end{acknowledgements}


\end{document}